\documentclass[aps,prl,onecolumn,showpacs,superscriptaddress]{revtex4}

\usepackage{amsfonts}
\usepackage{amsmath}
\usepackage{graphicx}        

\newcommand{\rme}{{\mathrm{e}}}

\newcommand{\rmpp}{{\mathrm{p}}}
\newcommand{\rmd}{{\mathrm{d}}}
\newcommand{\rmi}{{\mathrm{i}}}

\begin{document}

\title{Complexity and simplicity of plasmas}

\author{D. F. Escande}
  \affiliation{UMR 7345 CNRS-Aix-Marseille-Universit\'e, Facult\'e
de St J\'er\^ome, case 321, \\ Av. Normandie Niemen, FR-13397 Marseille CEDEX 20}

\begin{abstract}
This paper has two main parts. The first one presents a direct path from microscopic dynamics to Debye screening, Landau damping and collisional transport. It shows there is more simplicity in microscopic plasma physics than previously thought. The second part is more subjective. It describes some difficulties in facing plasma complexity in general, suggests an inquiry about the methods used empirically to tackle complex systems, discusses the teaching of plasma physics as a physics of complexity, and proposes new directions to face the inflation of information.
\end{abstract}


\maketitle

This paper is built upon a series of elements which aggregated thanks to the Senfest: my talk "Facing plasma complexity", the Panel Discussion on Nonlinear Dynamics and Complexity, and several interactions with the participants during the meeting and after. Plasma complexity is the thread unifying these elements, but their nature was heterogeneous. This is reflected in the organization of this paper. The first part is written like a classical scientific paper, and presents new results about microscopic plasma physics obtained directly from $N$-body dynamics: there is more simplicity in this physics than previously thought. The second part is more subjective, and describes some difficulties in facing plasma complexity in general, and possible directions to alleviate them, in particular in teaching.

\section{Direct path from microscopic dynamics to Debye screening, Landau damping and collisional transport}
\label{Dp}

When dealing with complex systems, it is generally admitted that one must give up the most fundamental descriptions of physics, and use more synthetic models; as Goldenfeld and Kadanoff \cite{GoKa} formulated it : ``Don't model bulldozers with quarks". In the case of plasma physics, its $N$-body description has been felt as formidable, and modelling microscopic plasma physics with kinetic equations has been considered as a necessity for decades now. However, a quite recent $N$-body approach \cite{DCb} figures Debye shielding and Landau damping in a very compact way. It also reveals that shielding and collisional transport are two aspects of the deflection of particles by Coulombian interactions. The introduction of a screened Coulomb potential for the $N$-body system then enables to tackle anew the issue of collisional transport, and to derive the first expression of transport coefficients incorporating all spatial scales, and in particular the inter-particle distance that was out of reach in traditional approaches.

This section deals with the One Component Plasma (OCP) model \cite{Salp,Abe,BH}, which considers the plasma as infinite with spatial periodicity $L$ in three orthogonal directions with coordinates $(x,y,z)$, and as made up of $N$ electrons in each elementary cube with volume $L^3$. Ions are present only as a uniform neutralizing background, which enables periodic boundary conditions. This choice is made to simplify the analysis which focuses on $\varphi(\textbf{r})$, the potential created by the $N$ particles at any point where there is no particle. The discrete Fourier transform of $\varphi$, readily obtained from the Poisson equation, is given by $\tilde{\varphi}(\textbf{0}) = 0$, and for $\textbf{m} \neq \textbf{0}$ by
\begin{equation}
  \tilde{\varphi}(\textbf{m})
  = -\frac{e}{\varepsilon_0 k_{\textbf{m}}^2} \sum_{j \in S}
     \exp(- \rmi \textbf{k}_{\textbf{m}} \cdot \textbf{r}_j),
\label{phitildetotM}
\end{equation}
where $e$ is the elementary charge, $\varepsilon_0$ is the vacuum permittivity, $\textbf{r}_j$ is the position of particle $j$, $S$ is the set of integers from 1 to $N$, $\tilde{\varphi}(\textbf{m}) = \int \varphi(\textbf{r}) \exp(- \rmi \textbf{k}_{\textbf{m}} \cdot \textbf{r}) \rmd^3 \textbf{r}$, with $\textbf{m} = (m_x,m_y,m_z)$ a vector with three integer components, $\textbf{k}_{\textbf{m}} = \frac{2 \pi}{L} \, \textbf{m}$, and $k_{\textbf{m}} = \|\textbf{k}_{\textbf{m}}\|$. Reciprocally,
\begin{equation}
\varphi(\textbf{r}) = \frac{1}{L^3}\sum_{\textbf{m}} \tilde{\varphi} (\textbf{m}) \exp(\rmi \textbf{k}_{\textbf{m}} \cdot \textbf{r}),
\label{phiInv}
\end{equation}
with the sum $\sum_{\textbf{m}}$ over all components of $\textbf{m}$ running from $- \infty$ to $+ \infty$.

The dynamics of particle $l$ is determined by Newton's equation
\begin{equation}
  \ddot{\textbf{r}}_l
  = \frac{e}{m_\rme} \nabla \varphi_l(\textbf{r}_l),
\label{rsectot}
\end{equation}
where $m_\rme$ is the electron mass, and $\varphi_l$ is the electrostatic potential acting on particle $l$, i.e.\ the one created by all other particles and by
the background charge. Its Fourier transform is given by equation (\ref{phitildetotM}) with the supplementary condition $j \neq l$.
Let $\textbf{r}_{l0}$  and $\textbf{v}_{l}$ be the initial position and velocity of particle $l$, and let $\delta \textbf{r}_l = \textbf{r}_l - \textbf{r}_{l0} - \textbf{v}_{l} t$. In the following, we consider the $\delta \textbf{r}_l$'s to be small. Therefore we approximate $\tilde{\varphi}_l(\textbf{m})$ by $\tilde{\phi}_l (\textbf{m})$, its expansion to first order in the $\delta \textbf{r}_l$'s. We further consider $\varphi$ to be small, and the $\delta \textbf{r}_l$'s to be of the order of $\varphi$. We now introduce the time Laplace transform which transforms a function $ f(t)$ into $\hat{f}(\omega) = \int_0^{\infty}  f(t) \exp(\rmi \omega t) \rmd t$ (with $\omega$ complex). We Laplace transform both the potential and the equations of motion. Combining the resulting equations yields
\begin{eqnarray}
 k_{\textbf{m}}^2\hat{\phi}(\textbf{m},\omega)
&-& \frac{e^2}{ L^3 m_e \varepsilon_0}
 \sum_{\textbf{n}} \textbf{k}_{\textbf{m}} \cdot \textbf{k}_{\textbf{n}}
\ \sum_{j \in S} \frac{\hat{\phi}(\textbf{n},\omega + \omega_{\textbf{n},j} - \omega_{\textbf{m},j})}{(\omega - \omega_{\textbf{m},j})^2} \exp[\rmi (\textbf{k}_{\textbf{n}}-\textbf{k}_{\textbf{m}}) \cdot \textbf{r}_{j0}]
\nonumber\\
= k_{\textbf{m}}^2 \hat{\phi}^{(0)}(\textbf{m},\omega) & &,
\label{phihat}
\end{eqnarray}
where carets indicate the Laplace transformed versions of the quantities, $\omega_{\textbf{l},j} = \textbf{k}_{\textbf{l}} \cdot \textbf{v}_{j}$, and $\hat{\phi}^{(0)}(\textbf{m},\omega) $ is the Laplace transform of $\tilde{\varphi} (\textbf{m}) $ computed by substituting $\textbf{r}_l$ by its ballistic approximation $\textbf{r}_{l0} + \textbf{v}_{l} t$ in Eq. (\ref{phitildetotM}). Equation (\ref{phihat}) is a linearized version of Eq. (15) of \cite{DCb}. It is the fundamental equation of this approach, and is of the type ${\mathcal{E}}\hat \phi=$ source term, where ${\mathcal{E}}$ is a linear operator, acting on the infinite dimensional array whose components are all the $\hat{\phi}(\textbf{m},\omega)$'s.

We introduce a smooth function $f(\textbf{r},\textbf{v})$, the smoothed velocity distribution function at $t=0$. We assume it to be a spatially uniform distribution function $f_0(\textbf{v})$ plus a small perturbation of the order of $\phi$. We replace the discrete sums over particles in Eq. (\ref{phihat}) by integrals over $f(\textbf{r},\textbf{v})$, and we keep the lowest order term in $\phi$. Then operator $\mathcal{E}$ becomes diagonal with respect to both $\textbf{m}$ and $\omega$, and Eq. (\ref{phihat}) becomes \cite{DCb}
\begin{equation}
  \epsilon(\textbf{m},\omega) \hat{\Phi}(\textbf{m},\omega)
  = \hat{\phi}^{(0)}(\textbf{m},\omega),
\label{phihatL}
\end{equation}
where $\Phi$ is the new approximation of $\phi$, and
\begin{equation}
  \epsilon(\textbf{m},\omega)
  = 1 - \frac{e^2}{L^3 m_\rme \varepsilon_0}
     \int \frac{f_0(\textbf{v}) }{(\omega -\textbf{k}_{\textbf{m}}  \cdot \textbf{v})^2} \ \rmd^3 \textbf{v}.
\label{eps}
\end{equation}
This shows that the smoothed self-consistent potential $\hat{\Phi}$ is determined by the response function $\epsilon(\textbf{m},\omega)$. The latter is the classical plasma dielectric function. A first check of this can be obtained for a cold colisionless plasma: then $\epsilon(\textbf{m},\omega) = 1 - {\omega_{\rmpp}^2}/{\omega^2}$,
where $\omega_{\rmpp} = [(e^2 n)/(m_\rme \epsilon_0)]^{1/2}$ is the plasma frequency ($n = N/L^3$ is the plasma density). The classical expression involving the gradient of $f_0$ in $\textbf{v}$ is obtained by a mere integration by parts.

By inverse Fourier-Laplace transform, to lowest order the contribution of particle $j$ to $\hat{\phi}^{(0)}(\textbf{m},\omega)$ turns out to be, after some transient, the \textit{shielded Coulomb potential} of particle $j$ \cite{Gasio,Bal,Rost}
\begin{equation}
  \delta \Phi_j (\textbf{r})
  = \delta \Phi(\textbf{r} - \textbf{r}_{j0} - \textbf{v}_j t,\textbf{v}_j),
\label{phij}
\end{equation}
where
\begin{equation}
  \delta \Phi (\textbf{r},\textbf{v})
  = - \frac{e}{L^3 \varepsilon_0} \sum_{{\textbf{m}} \neq {\textbf{0}}}
      \frac{\exp(\rmi \textbf{k}_{\textbf{m}} \cdot \textbf{r})}
           { k_{\textbf{m}}^2 \, \epsilon(\textbf{m},\textbf{k}_{\textbf{m}} \cdot \textbf{v})}.
\label{phi}
\end{equation}
Therefore, after this transient, the dominant contribution to the full potential in the plasma turns out to be the sum of the shielded Coulomb potentials of individual particles located at their ballistic positions.

We now apply the smoothing using distribution function $f$ to $\tilde{\phi}^{(0)}(\textbf{m},\omega)$ too in equation (\ref{phihatL}). To lowest order in the $\delta {\textbf{r}}_{j}$'s, this approximates $\hat{\Phi}(\textbf{m},\omega)$ by
\begin{equation}
  \hat{\Phi}^{(0)}(\textbf{m},\omega)
  = - \frac{\rmi e}{\varepsilon_0 k_{\textbf{m}}^2} \int
    \frac{\tilde{f}(\textbf{m},\textbf{v})}
         {\omega -\textbf{k}_{\textbf{m}} \cdot \textbf{v}} \
     \rmd^3 \textbf{v}.
\label{phi0hatcg}
\end{equation}
This shows that this second smoothing makes equation (\ref{phihatL}) to become the \textit{expression including initial conditions in Landau contour calculations of Langmuir wave growth or damping}, usually obtained by linearizing Vlasov equation and using Fourier-Laplace transform, as described in many textbooks. Therefore, in these calculations, $\hat{\Phi}(\textbf{m},\omega)$ turns out to be the smoothed version of the actual shielded potential in the plasma.

Picard iteration technique applied to  Eq. (\ref{rsectot}) enables to get insight into the \textit{nature of Debye shielding} \cite{DCb}, and yields the following intuitive picture. Consider a set of randomly distributed particles at $t =0$. Consider a particle $l$. At a later time, it has deflected all particles which made a closest approach to it. Therefore the number of particles in a sphere about particle $l$ is smaller than when the Coulomb repulsion is absent. This means that the effective charge of particle $l$, as seen out of the sphere, is reduced due to this repulsion: the charge of particle $l$ is shielded due to the deflections. If a larger sphere is considered, more particles are deflected, and the shielding is larger: the shielding of particle $l$ increases when the distance from this particle increases. Shielding is almost complete at a distance of the order of the Debye length. It is completed after the time necessary for a thermal particle to cross this distance, i. e. the plasma period. This fixes the duration of the above transient. Consequently, \textit{shielding is a cooperative dynamical process: it results from the accumulation of almost independent repulsive deflections with the same qualitative impact on the effective electric field of particle} $l$ (if ions were added, the attractive deflection of charges with opposite signs would have the same effect). It is a cooperative effect, but not a collective one (it does not involve any synchronized motion of particles). Basic plasma physics textbooks show the accumulation of almost independent repulsive deflections to produce collisional transport of particles in plasmas. Unexpectedly, it turns out that \textit{Debye shielding is another aspect of the same two-body repulsive process.}

In order to \textit{describe collisional transport}, we replace the original $N$-body problem by one where the screened Coulomb potential is substituted for the bare Coulomb one. Then the deflection of particle $l$ can be performed in a sequence of steps \cite{DCb}. First, we use first order perturbation theory in $\delta \Phi$ as given by Eq.~(\ref{phi}), which shows the total deflection to be the sum of the individual deflections due to all other particles. For an impact parameter much smaller than $\lambda_{\rm{D}}$, the deflection due to a particle turns out to be the perturbative value of the Rutherford deflection due to this particle if it were alone. Second, for a close encounter with particle $n$, we show that the deflection of particle $l$ is exactly the one it would undergo if the other $N-2$ particles were absent. Third, the deflection for an impact parameter of order $\lambda_{\rm{D}}$ is given by the Rutherford expression multiplied by some function of the impact parameter reflecting shielding. These three steps yield an analytical expression for deflection whatever the impact parameter. When $\textbf{v}_l$ is small, $\delta \Phi (\textbf{r},\textbf{v}) \simeq \delta \Phi (\textbf{r},\textbf{0})$ which is the Yukawa potential $\delta \Phi_{\rm{Y}} (\textbf{r}) = - \frac{e}{4 \pi \varepsilon_0 \| \textbf{r} \|} \exp (- \frac{\| \textbf{r} \|}{\lambda_{\rm{D}}})$ (Eq.~(18) of Ref.\ \cite{Gasio}). The first order correction in $\textbf{k}_{\textbf{m}}  \cdot \textbf{v}_l$ to this approximation is a dipolar potential with an electric dipole moment proportional to $\textbf{v}_l$. Since a Maxwellian distribution is symmetrical in $\textbf{v}$, these individual dipolar contributions cancel globally. As a result, the first relevant correction to the Yukawa potential is of second order in $\textbf{k}_{\textbf{m}}  \cdot \textbf{v}_l$. This should make the Yukawa approximation relevant for a large part of the bulk of the Maxwellian distribution. With this Yukawa approximation, the calculation can be pursued analytically. One finds that the classical Coulomb logarithm $\ln (\lambda_{\rm{D}} / \lambda_{\rm{ma}})$ of the second Eq.~(14) of Ref.\ \cite{Ros} giving the collisional diffusion coefficient becomes $\ln (\lambda_{\rm{D}} / \lambda_{\rm{ma}}) + C$ where $C$ is of order unity (here $\lambda_{\rm{ma}}$ is the distance of minimum approach of two electrons in a Rutherford collision, as given by energy conservation). If the full dependence of the shielding on $\textbf{v}_l$ were taken into account, the modification of the Coulomb logarithm would be velocity dependent.

As a conclusion of this section, it is interesting to put in perspective its approach with that of classical plasma physics textbooks. In the present approach, the smoothed velocity distribution is introduced after particle dynamics has been taken into account, and not before, as occurs when kinetic equations are used (see for instance chapter III of \cite{LiPi}). The former approach avoids addressing the issues of the exact definition of the smoothed distribution for a given realization of the plasma, and of the uncertainty as to the way the smoothed dynamics departs from the actual $N$-body one, as explained in section \ref{Ddpc}. Readers familiar with the Klimontovich equation might feel a resonance with the present $N$-body approach. Indeed this equation describes the evolution of the granular distribution function corresponding to the exact $N$-body dynamics. However, it is so intricate that, as yet, it has required approximate versions to be introduced before performing any concrete calculations about the distribution function dynamics.

An unexpected outcome of the present approach is the passage through the derivation of the Debye shielded potential of the $N$-body system in order to derive the setting for the classical calculation of Landau damping. Classical plasma physics textbooks introduce Debye shielding in a completely independent way. They often appeal to the ability of particles to move and neutralize any region of excess space charge, which makes sense if there is a macroscopic polarized Langmuir probe, but not for uniform plasmas. Sometimes they single out a test particle, and find by kinetic calculations that it is shielded by all the other ones, but this does not reveal how all particles shield the other ones and are also shielded by them at the same time. The present approach uncovers that shielding is a mere consequence of the independent deflections of particles due to the Coulomb force.

The discussion of collisional transport in classical plasma physics textbooks (see for instance chapter IV of \cite{LiPi}) uses the two-body picture and the Balescu-Lenard picture. However none correctly takes into account the interaction of particles at distances about the inter-particle distance. Indeed, the Balescu-Lenard approach cannot describe the graininess of these scales, and the two-body picture cannot describe the simultaneous ``collisions" with several particles. The calculation of collisional transport described in this section, and more thoroughly in \cite{DCb}, takes into account the interactions at all distances for the first time.

\section{Facing plasma complexity}
\label{Fpc}
This issue was already discussed at length in section II of the paper ``How to face the complexity of plasmas?" \cite{Howto}. This paper proposed the following definition of ``complexity", a word with a lot of meanings. The one used in this section is ubiquitous in modern science. However, this meaning comes with an increasing number of attributes when going from inanimate matter to living matter, to humans, and to societies. These attributes may be ordered in a sequence of levels of description. For inanimate matter, the number of levels is smaller than for living systems. The coarsest level has two aspects: on the one hand, the whole is more than its parts and on the second hand it displays a spontaneous self-organization. The former aspect may be very strong: the whole may be a lot beyond the sum of its parts, as occurs for open systems, be it a plasma column in a laboratory, or the human body whose matter is almost completely renewed about every two months through metabolism and repair. The latter aspect is important to tell complex systems from artifacts like computers or engines. It may come with two opposite, but possibly interrelated, features: order and chaos.

Emergence is the central feature of the level of description following the largest one. It results from self-organization, and is the appearance in the system of interest of a feature (form or pattern) arising out of a multiplicity of relatively simple interactions of smaller parts. This feature cannot be anticipated from the knowledge of the parts of the system alone, even whenever these parts are also complex systems made up of finer scales. A typical example of emergent structure is a fluid vortex, as occurring for instance due to the motion of water particles in a pipe flow. In turn, individual vortices may interact to produce another emergent feature: turbulence. This emergence makes the water less fluid than in the laminar state: pressure drop increases, but molecules are unaffected; again, the whole is a lot beyond the sum of its parts. Debye shielding is an emergent feature of the OCP model, since it is not an obvious consequence of the simple rules defining the model.

In living systems, next levels of description include features like cooperation and competition. There are other definitions of complexity, which introduce the same attributes in a different order, or other attributes, which are implicitly present in the above definition: complex systems contain many interdependent constituents interacting nonlinearly, and their self-organization spans several spatial and temporal scales.

We now describe some difficulties in the description of plasma complexity, we suggest an inquiry about the methods used empirically to tackle complex systems, and we suggest to teach plasmas as complex systems.

\subsection{Difficulties in the description of plasma complexity}
\label{Ddpc}

Section II.A of \cite{Howto} already discussed this issue at length. The next paragraph summarizes some of the difficulties listed therein, before expanding the discussion in a direction motivated by the first section of this chapter.

Very often, \textit{models used in plasma physics have feet of clay}, since they cannot be derived axiomatically from first principles with conditions of validity suited to their actual applications: fluid models are not justified from first principles for collisionless plasmas, and the validity of solutions of Vlasov equation is proved only up to the time of exponential divergence of nearby orbits  \cite{Spohn}, a too stringent time bound to describe turbulence. Each physicist must elaborate his/her own global view about plasma physics from many \textit{models which do not have any strict hierarchy}; indeed any model may be complexified in many ways: by increasing the dimensionality, the complexity of geometry, the number of involved physical quantities, by going from a fluid to a kinetic description, etc... A \textit{principle of simplicity} (Occam's razor principle) dominates the modeling activity. The \textit{validation of assumptions} turns out to be more \textit{difficult} for a complex system than for a simple one, because of the limited amount of information about it. At any moment, the description of plasma complexity is provisional, and results from a collective and somewhat unconscious process. All this makes changing views more difficult.

Though often considered as the most accurate way to describe plasma dynamics, kinetic equations come with difficult issues: what is the exact definition of a smoothed distribution function of velocities for a given realization of the plasma? How long does the kinetic solution stay close to the actual granular plasma dynamics? Even tougher is the issue of the analytical tools available to describe nonlinear dynamics and chaos for partial differential equations (PDE's). This latter point was the incentive to tackle the weak warm beam-plasma instability by generalizing \cite{Esc91,Ten} a model originally introduced for the numerical simulation of the cold beam-plasma instability \cite{OLMSS,ONWM}. Indeed, there are more tools for nonlinear dynamics and chaos available for ordinary differential equation (ODE's) than for PDE's.

In general, the $N$-body description of plasma physics problems is formidable, even for closed systems. The actual difficulty is even higher, since most systems considered in plasma physics are open systems: the mechanical system to describe is elusive! Kinetic equations or fluid models look more adapted since their ODE structure is compatible with source and sink terms, and with boundary conditions. However, they are generally derived for closed systems and modified in an ad hoc way to tackle the open ones. Furthermore, their use requires the assignment of boundary conditions, a mathematical requirement which may be in conflict with the physics process of interest. For instance, the boundary values of the density in a tokamak results from the self-organization due to the transport of particles and of the plasma-wall interaction, while its description by the convection-diffusion model (or Fokker-Planck equation), requires explicit boundary conditions to solve the differential equation. Such a difficulty is avoided if one describes transport with an integro-differential equation like the Chapman-Kolmogorov one, or if the wall is included in a global plasma-wall dynamics.

\subsection{Toward a methodology for complexity ?}
\label{Tmc}

There is no science of complexity, but we might benefit from finding out the common aspects of the methods used empirically to tackle complex systems. Possibly a methodology might emerge from this empirical knowledge accumulated over a wealth of research works. We first discuss some topics which may be considered in this inquiry, and then illustrate a particular case: how several aspects of complexity in biology are suggestive for analyzing that of fusion plasmas.

A first topic is the description of complex systems by models. Since such systems are far from elementary physics, they are described by a multiplicity of models built from first principles, like conservation laws, or by a derivation from more fundamental equations, by some stochastic Ansatz, etc...
How are embedded these models? How are they validated? In plasma physics fluid equations can be derived \`a la Chapman-Enskog or by appropriate closures of kinetic equations, but are often used without making sure that the corresponding approximations are justified. Then conservation laws may be invoked, but the eventual justification of the model is its ability to make relevant interpretations of experiments, or to make predictions, and, even better, experimentally verified ones.

Then comes the issue of the way these models are used: for analytic calculations, for numerical ones? What is the link between these approaches? The complexity described by numerical simulations is much smaller than in actual experiments. Simulations often come without error bars on their predictions, the role of intentionality is higher therein than in analytical calculations because of the choice of initial conditions and of parameters, and the numerical coding of an analytical model often involves many uncontrolled approximations. For instance, rigorously speaking, Vlasov codes are non Vlasovian: the ``fluidity" of phase space is lost in PIC codes at the expense of a numerical noise, while codes keeping this fluidity smear out the fine Vlasovian structures of phase space. Analytical calculations are more limited than numerical ones, but they reveal the internal structure of a model and its dominant parameters; they have an intrinsic flexibility with respect to the parameters values, and they can be checked more easily than a numerical simulation. Analytical calculations can be used to verify numerical simulations and avoid some of their pitfalls: cancelations of large terms, problems of stability and convergence due to insufficient numerical analysis, fake boundary effects, fake dissipation, etc...

Another topic is the type of emergence (large form or pattern due to smaller parts) which is present in the complex system of interest. How is it described theoretically? Does it induce the derivation of new descriptions? How do order and chaos coexist? Processes which appear complicated at one scale often produce emergent features in a suitable limit at a higher scale that are relatively simple \cite{ASS}. For instance, the Taylor-Green vortex can be easily built from the hydrodynamic description of the fluid provided by the incompressible Navier-Stokes equation, and not from its $N$-body description. Finally reviewing present theories of self-organization might prove useful too.

Physicists might \textit{get inspiration from the complexity of living beings}. Indeed such ``systems" exhibit aspects of complexity which may be overlooked in the complex systems of physics. They involve many cooperating processes of self-organization, which induce emerging features including the living being itself at the largest scale. Biologists enumerate their feedback loops, and find out the relevant ones for a given emergent feature. Such an analysis might prove useful for magnetic confinement devices where many self-organization processes are at work (see the feedback loops in figure 5 of \cite{ITER2Trsp}) inside the plasma, at the plasma edge, but also involving the machine itself. With such a picture explicitly in view, the analysis of experimental results might start on a firmer basis. For instance, there has been since 1995 the puzzle of the cold-pulse experiments exhibiting a fast transient increase in the electron temperature in the plasma core in response to an abrupt cooling of the edge (see \cite{Pust} for an exhaustive bibliography). It is only quite recently that the induction effects and the fast energy exchange between the plasma and the magnetic field were combined with the traditional transport model to tackle this issue \cite{Pust}. Such an approach might have been studied much faster with a more global view of the complex system under study.

Another striking aspect of living beings is that, in a loose sense, they are their own cause. Something of this kind already exists in the dynamo of the nonlinear tearing mode, or of the reversed field pinch (RFP; see section 2 of \cite{CapIAEA}). Then the helical deformation of the system which originates in a magnetic instability comes with a flow produced by an electrostatic drift which sustains the structure. Self-sustainment is also at work in the transport barriers of magnetic confinement devices. In this case one might even think of such barriers as ``membranes". Indeed, the part of the system inside the barrier has a proper dynamics which is regulated by exchanges through the ``membrane" with the external part. Though of paramount importance, a more hidden membrane is the region of plasma-wall interaction.

In biology, Darwinian selection plays an important role. Something of this kind is also at work in fusion plasmas. For instance many MHD modes may compete, but whenever one wins, the helical deformation it imposes may kill all the other ones, and lead to a regular structure: this is the case for single helicity states in RFP's.

Biologists produce global pictures of the exchange of matter with the environment, of the energy consumption, and of the production of heat and motion in living beings. As yet, this has been a lot less explicit for fusion devices. For instance fueling is an important topic for experimentalists, but it has attracted little theoretical attention, in particular as far as its connection with heat transport is concerned.

Biological metabolism is a vast autocatalytic system, in particular because of the enzymatic process, which accelerates chemical reactions by several orders of magnitude, and therefore energy consumption too. Something of this kind is also at work in fusion plasmas. Indeed, anomalous transport is a way for fusion systems to increase considerably their energy consumption for given thermonuclear conditions. Furthermore, when the system develops helical structures, like magnetic islands, viscous dissipation is added to the ohmic one because of the dynamo at work.

Cooperation and competition are important aspects of living systems. Competition of MHD modes was discussed earlier. It also occurs to produce the intermittent back transitions to multiple helicity in RFP's. Cooperation is at work in anomalous transport, which results of the compound effect of a series of modes resonating at various radial positions.

\subsection{Teaching plasma complexity}
\label{Tpc}

Teaching plasma physics as physics of a complex system would prepare students to the genuine nature of plasmas, and would lead teachers to face more explicitly plasma complexity. To this end, complexity should be defined... with the caveat that there is no single definition of this word in a scientific context. One should warn the students from the outset:

-   Most experimental and natural plasmas display self-organization and emergence.

-   There is a multiplicity of models, which are not necessarily proved from first principle to be correct for the type of plasma of interest.

-   Teaching must start from simple cases, but be prepared to deal with more complex ones that correspond to real plasmas.

Difficulties in the description of plasma complexity mentioned in section \ref{Ddpc} might be indicated as well.

The various chapters of a plasma course might be put in a broader perspective. For instance, after dealing with waves in homogeneous plasmas with fluid models, one might indicate what occurs when there are gradients of the main plasma parameters, or when kinetic effects are included.

Intuitive explanations are powerful memory anchors. To provide them, it may be useful to separate concepts from mathematical tools. For instance, the Fourier decomposition sometimes hides the nature of the physics underlying waves: the plasma frequency is naturally introduced as that of a harmonic oscillator corresponding to the vibration of an electron slab with respect to its neutralizing ion slab. Langmuir waves are then understood by placing side by side such slabs where nearby ones have electrons in phase opposition (see section 14.2.2 of \cite{Houches}); similarly, drift waves can be intuitively understood as the juxtaposition of the drift bumps described intuitively in section II.A of \cite{Horton}. The mechanical approach of section \ref{Dp} may be useful from this point of view, since it enables to introduce Debye shielding and Landau damping without having to use kinetic equations.

Basic plasma textbooks are full with linear physics, while nonlinearity rules plasma complexity. It might be appropriate to shift the balance toward the latter aspect. Also to warn students that the smallness of a perturbation is not a sufficient criterion to validate linear calculations. Indeed, as reviewed in Ref.\ \cite{HS}, perturbation theory that relies on linearization has to be questioned, as it yields a solution of the linearized set of equations only. Whether it also generates a solution of the full set has to be shown explicitly, and this may be a hard task. For instance the full proof of existence of Landau damping \cite{MV} in a Vlasovian frame was a mathematical tour de force, the equivalent of a Kolmogorov-Arnold-Moser theorem for continuous systems, and led its main author to be awarded the 2010 Fields medal. Finally, nonlinear dynamics and chaos might provide a way to revisit and unify often separated chapters, e.g.~turbulent and collisional transport, the calculation of magnetic field lines, or the introduction of fluid and of Vlasov equations.

\subsection{Facing the publication deluge}
\label{Fpd}

The complexity of plasmas comes with inflation of topics and of information to cope with. To this end, several methodological improvements were proposed in section II.B of \cite{Howto}. Here we recall three of them: improving the ways papers are structured, improving the way scientific quality is assessed in the referral process, and developing new databases.

The suggested improvement to the structure of papers is the following: each paper, even letters, would have a ``\textit{claim section}" being a kind of executive summary. It would summarize the main results and their most relevant connection to previous literature, making this information easy to get without entering the arcanes of the paper. It would provide a clear information about the importance, the originality, the actual scientific contribution of the paper, and about the ``precedents, sources, and context of the reported work" as worded in the APS guidelines for professional conduct. Salient figures or formulas would be set there to support the claims. This procedure would suppress the "can't see the forest for the trees" syndrome. Indeed, it is often difficult to get the point of a given paper from its abstract, introduction and conclusion; in particular to find out what are the most important figures or formulas. The very compact way a claim section would communicate new results is reminiscent of the way physicists communicate the essence of their latest paper to their colleagues in the corridor of large congresses. It would also provide a reader with what he tries to get by himself when looking for the essence of a paper without reading it entirely. It would also improve the clarity of the papers by driving an author to state the essence of his/her results in a more accessible way, and without having to care about the literary constraints of a normal text. Finally one may notice that this procedure is the classical one in mathematical papers, and that the ``highlights" in Physics Letters A go somewhat in this direction.

With this tool, the \textit{referral process might be improved} by requiring referees to check the claims of the claim section, and to motivate their possible disagreements with any of them. This procedure should make the referral process more scientifically rigorous, more ethical, and faster. Editors would benefit from a better refereeing process, which would avoid many authors' complaints, while accelerating the editorial process. Journals would benefit from the increased clarity of the contents of their published papers. The procedure might start with an experimental stage where the claim section would be optional for the authors, but not for the referees if the claim section is available.

The claim sections might be set by each scientific journals or publisher into a new \textit{dedicated database} accessible through Internet where cross-referenced claim sections would be hyperlinked. This would provide a more focused technique for data retrieval adapted to plasma complexity. It would ease the assessment of the state of the art of a given topic, with respect to what is available through present bibliographical databases. Indeed all the references of a published paper do not have the same importance, but their equal treatment in present databases generates a huge amount of information to be screened when going forward or backward in time, while following the quotation links.

Finally, it might be useful to go further in the direction inaugurated by the ITER Physics Data Base. International groups of experts gathered aiming at a consensus about the various issues of tokamak physics, and already published two thick issues of Nuclear Fusion summarizing their views \cite{ITER,ITER2}. Such a process might be enlarged to magnetic fusion in general, and even better, to the whole physics of plasmas. Indeed, the generation of pioneers of this physics is now fading away, which rules out the dream of having physicists with all of it in their mind. Therefore, for the younger generation of physicists it would be extremely useful to benefit from periodically updated collective reviews providing an in-depth presentation of plasma physics.

\section{Conclusion}
\label{Concl}
The first part of this chapter introduced Debye shielding, Landau damping, and collisional transport through a mechanical approach. This approach brings unification and simplification into basic microscopic plasma physics, and may be useful for pedagogical purposes. It also provides the first calculation of collisional transport accounting correctly for all impact parameters. One might think about trying to apply this approach to plasmas with more species, or with a magnetic field, or where particles experience trapping and chaotic dynamics. The first generalization sounds rather trivial. The third one is on its way, at least in one dimension, by using a Hamiltonian describing wave-particle interaction \cite{Esc91,Ten} (see a pedestrian introduction in \cite{Houches} and more specific results in \cite{BEEB,BEEBEPS}). Here we introduced only the linearized version of the fundamental nonlinear equation (15) of \cite{DCb}. The nonlinear version opens interesting prospects, like the study of the effect of the coupling of Fourier components with both coherent and incoherent aspects.

The description of complexity of plasmas is an intricate issue, and the future development of the corresponding methodology is even more intricate. Collective effects are important in plasma physics, but also for its development: it requires collective efforts of plasma physicists. This chapter would like to contribute to a brainstorming in the plasma community. Indeed, it would be very useful for this community to pay attention to the essence of its physics and of its practice. To this end it should review past published material, but also its past way of thinking, of interacting, and of meeting together.

As to meetings, it is striking to see in large conferences the ever increasing number of people physically present at a talk, but actually busy with some electronic device. The speaker is far away, often confined in a precise spot, and his/her picture is visible on a large screen. This might indicate that traditional conferences are becoming obsolete, and that they should be made more interesting by decreasing the time where participants are passive. For instance, invited talks might be recorded in advance, and available on-line before the meeting. The meeting would be shorter, keeping its poster sessions, and concentrating the oral sessions on the discussion of the various recorded talks with more time for this purpose than the traditional ``five minutes". Another possibility would be to devote more time to review talks or panel discussions updating the information about topics with a broad scope.

I thank the Organizing Committee of ICPPNDS for inviting me, A. Das for her help, and A. Surjalal Sharma for pointing references \cite{GoKa} and \cite{ASS} out to me. L. Cou\"edel, R. Ganesh, and F. Sattin are thanked for making useful comments to a first draft.

\end{document}